\documentclass{ckm}                 
\usepackage{txfonts}            

\confname{Workshop on the CKM Unitarity Triangle, IPPP Durham, April
  2003}

\title{Physics potential of the decays 
$\mathrm{B_{s,d} \rightarrow J/\psi\eta}$ and 
$\mathrm{B_{s} \rightarrow J/\psi\phi}$ 
in the ATLAS experiment at the LHC}

\author{C Driouichi}
\address{Lund University Sweden, for the ATLAS collaboration}

\begin{document}

\begin{abstract} The CP asymmetry predicted by the Standard Model in the 
decay modes $\mathrm{B_{s}\rightarrow J/\psi+\eta}$ and 
$\mathrm{B_{s}\rightarrow J/\psi+\phi}$ is very small, and the observation of 
a sizeable effect would be a clear indication of new physics beyond the 
Standard Model. Contrary to the final state $\mathrm{J/\psi\phi}$, the former 
final state is a CP eigenstate, and therefore complex angular distribution 
analyses are not necessary for the asymmetry measurement. In this paper, the 
approaches for the study of these two decay channels are presented.  
\end{abstract}

\maketitle

\section{Introduction}

The decay mode $\mathrm{B_{s} \rightarrow J/\psi\eta}$ is analogous to the 
mode $\mathrm{B_{s} \rightarrow J/\psi\phi}$, which has been studied 
extensively in view of CP violation measurements.
In these two decay modes, the CP asymmetry predicted by the Standard Model is 
very small, and the observation of a sizeable effect would be a clear signal 
of physics beyond the Standard Model. 

The decay channel $\mathrm{B_{s} \rightarrow J/\psi\eta}$ can be used to 
measure 
various parameters in the $\mathrm{B_{s}}$-meson system. 
In the ATLAS experiment, assuming an integrated luminosity of 30~fb$^{-1}$, 
roughly 10 000 decays can be reconstructed with a signal-to-background ratio 
of about 1:1, as will be shown later in this paper. It can be thus foreseen 
that the measurement of the branching fraction of this mode, the measurement 
of the $\mathrm{B_{s}}$ lifetime, as well as other measurements will be 
feasible, which constitues an important cross-check of other measurements 
performed using other decay channels. 

Moreover, the decay mode $\mathrm{B_{s} \rightarrow J/\psi\eta}$, when 
combined with its U-spin symmetric channel, 
$\mathrm{B_{d} \rightarrow J/\psi\eta}$, provides a new strategy for the 
extraction of the $\gamma$ angle of the Unitarity Triangle~\cite{skands}. 
Indeed, while the phase $\exp(i\gamma)$ is CKM suppressed for the decay mode 
$\mathrm{B_{s} \rightarrow J/\psi\eta}$, this is not the case for the 
decay mode $\mathrm{B_{d} \rightarrow J/\psi\eta}$. CP violation effects 
could thus potentially be more easily accessible through the latter mode by 
measuring the time-dependent asymmetry of $\mathrm{B_d}$. 
The overall normalization can be fixed by measuring CP averaged rates of 
$\mathrm{B_{d} \rightarrow J/\psi\eta}$ and 
$\mathrm{B_{s} \rightarrow J/\psi\eta}$, which assumes the validity of 
the U-spin symmetry (the approximate symmetry of u, d and s quarks). 
The measurements seem to be, however, out of reach due to the small expected 
branching ratio of $\mathrm{B_{d} \rightarrow J/\psi\eta}$, and the 
overlapping mass peaks of the $\mathrm{B_{s}}$ and 
$\mathrm{B_{d}}$~\cite{chafik}.

The $\mathrm{B_s-\bar{B}_s}$ system is characterized by two eigenstates 
with different masses and decay rates. The experimental determination of 
the mass difference $\mathrm{\Delta m_{s}}$ and the rate difference 
$\mathrm{\Delta \Gamma_{s}}$ will be valuable input for flavour dynamics 
in both the Standard Model and its possible extensions. While the 
measurement of $\mathrm{\Delta m_{s}}$ has been proven to be accessible 
through the decays $\mathrm{B_{s} \rightarrow D_{s}\pi}$ and 
$\mathrm{B_{s} \rightarrow D_{s} a_{1}}$, $\mathrm{\Delta \Gamma_{s}}$ is
expected to be precisely measured in ATLAS through the decay channel 
$\mathrm{B_{s} \rightarrow J/\psi\phi}$. 

\section{The $\mathrm{B_{s,d} \rightarrow J/\psi\eta}$ decay channel}

Signal samples of $\mathrm{B_{s,d} \rightarrow J/\psi\eta}$ were 
generated using the Monte Carlo program PYTHIA 5.7 \cite{pythia}. The 
J/$\psi$ meson was forced to decay into $\mu^+\mu^-$, and only events 
passing the ATLAS level-1 trigger requirements for B hadrons~\cite{tdr1} 
(a muon with a $p_{\rm T}>$~6 GeV and $|\eta| < 2.4$)\footnote{Throughout this paper, the 
symbol $p_{\rm T}$ is used for the transverse momentum with respect to the 
beam direction, and $\eta$ for the pseudorapidity.}
and the $\eta$ meson decaying into $\gamma\gamma$ were retained. The 
branching ratio for the decay $\eta \rightarrow \gamma\gamma$ is 
($39.33 \pm 0.25$)~\%\cite{pdg}. Events were further selected to satisfy 
the ATLAS second level trigger: the presence of a second muon 
with $p_{\rm T}>3$~GeV and $|\eta| < 2.5$. The resulting signal samples 
were consisting of about 17,000 ${\rm{B_{s}}}$ events and 15,000 
${\rm{B_{d}}}$ events.

A full GEANT-based simulation was used to simulate the response of the ATLAS 
inner detector and the electromagnetic calorimeter. The muon chambers were not 
included in the simulation, but only real muons were included in the analysis. 
The muon identification efficiencies were assumed to be 85\% for the first 
muon (the muon with $p_{\rm T}>$~6 GeV) and 78\% for the second muon 
(the muon with $p_{\rm T}>$~3 GeV). The 85\% efficiency for the muon with the 
higher transverse momentum includes both reconstruction and level-1 trigger 
efficiencies. 

The J/$\psi$ was reconstructed by fitting pairs of opposite-charge 
muons passing the level-2 selection ($\mu6\mu3$)\footnote{In the following, the 
notation $\mu6$ ($\mu3$) is used to indicate a muon with a $p_{\rm T}>$~6 (3) 
GeV.} to a common vertex and by calculating their invariant mass. The 
$\chi^2$/d.o.f. of the vertex fit was required to be less than 6.0, 
and the decay vertex of the J/$\psi$ was required to be detached from the 
primary vertex by at least 250~$\mu$m in the transverse plane. 
The J/$\psi$ mass resolution was found to be 39 MeV. The resolution of the 
J/$\psi$ decay vertex in the transverse plane was 64~$\mu$m~\cite{tdr2}. The 
efficiency of the reconstruction within a three-standard-deviation mass window 
was 79\%. If a J/$\psi$ was found, the analysis proceeded with 
$\eta$-reconstruction in the electromagnetic calorimeter. The $\eta$ 
reconstruction is described in more detail in Ref.~\cite{fairouz}. 
The mass distribution, Fig.~1, was fitted with two 
Gaussians, and the obtained mass resolution was $\mathrm{\sigma = 70~MeV}$.
The total reconstruction efficiecny was 2.3$\%$ within two standard 
deviations from the nominal mass.

\begin{figure}
\hbox to\hsize{\hss
\includegraphics[width=\hsize]{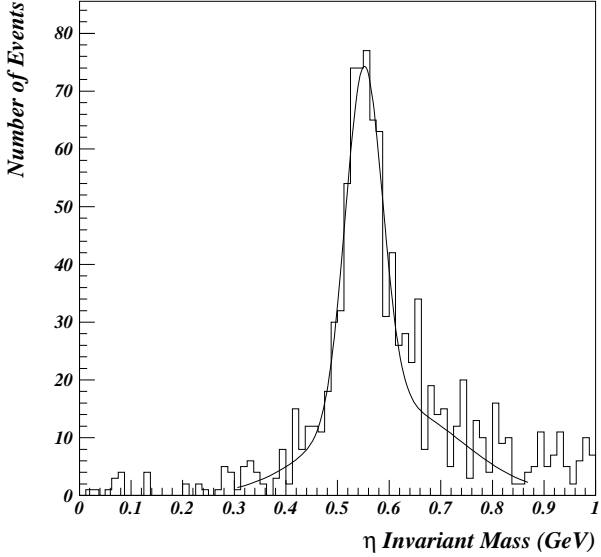}
\hss}
\caption{Invariant mass distribution of $\eta$-mesons in signal events after all the cuts.}
\label{fig:fig-eta}
\end{figure}

The J/$\psi$ and $\eta$ candidates were combined to reconstruct both the 
$\mathrm{{B_{s}}}$ and $\mathrm{{B_{d}}}$. A mass resolution of around 67 MeV 
was found for both channels, with an overall signal efficiency of 
$\sim$ 1$\%$ within three standard deviations from the nominal mass.

\begin{figure}
\hbox to\hsize{\hss
\includegraphics[width=\hsize]{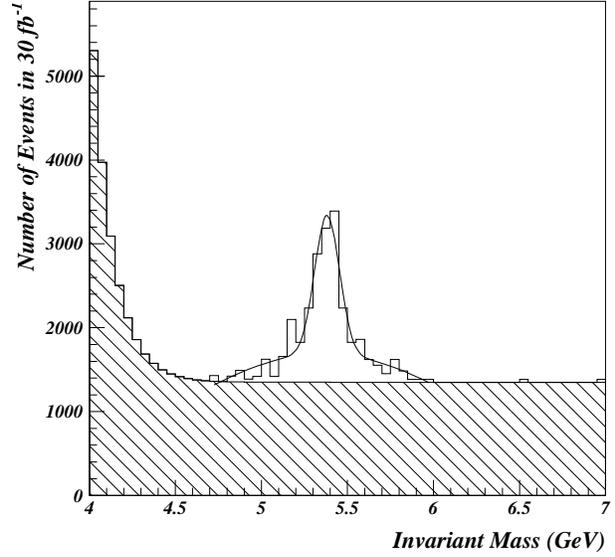}
\hss}
\caption{Invariant mass distribution of the reconstructed 
$\mathrm{B_s}$-signal (white) and the background (dashed).}
\label{fig:fig-sb}
\end{figure}

The background considered was a sample of $\mathrm{B \rightarrow J/\psi X}$ 
decays, processed and selected in the same way as the signal events. The 
resulting efficiency was less than 0.009$\%$. 
The table below summarizes the event rates for both signals and background.

\begin{table}[hbt]
\begin{center}
\begin{tabular}{|l|c|c|}     \hline
   &  $\theta_P = -10^o$              & $\theta_P = -20^o$ \\   \hline 
$\mathrm { \mathcal{B}r (B_s \rightarrow J/\psi\eta) }$ &  $8.3 \cdot 10^{-4}$    &   $9.5 \cdot 10^{-4}$   \\ 
$N\mathrm{_{Bs}^{obs}}$ & 8 400 & 9 600 \\ \hline
$\mathrm { \mathcal{B}r (B_d \rightarrow J/\psi\eta) }$ &  $4.1 \cdot 10^{-6}$    &   $1.6 \cdot 10^{-6}$   \\ 
$N\mathrm{_{Bd}^{obs}}$ & 200 & 80 \\ \hline
$N\mathrm{_{back}^{obs}}$ & 10 800 & 10 800 \\ \hline
\end{tabular}
\end{center}
\caption{Assumed branching ratios for $\mathrm{B_{s,d} \rightarrow J/\psi\eta}$, 
and the estimated numbers of signal and background events for an 
integrated luminosity of 30~fb$^{-1}$. $\theta_P$ is the $\eta-\eta'$ 
mixing angle ( see Ref.~\cite{etamixing}).}
\label{tab:br}
\end{table}

For the measurement of the CP asymmetry in the $\mathrm{B_s}$-system, the 
so-called Same-side jet-charge tagging method was used for distinguishing 
production of $\mathrm{B_s}$ and $\mathrm{\bar{B}_s}$ ~\cite{tdr2}. 

The jet charge was defined as :
\begin{eqnarray*}
Q_{\mathrm{jet}} = \frac{\sum_{i} q_i p_i^k }{\sum_{i} p_i^k },
\end{eqnarray*}
where $q_i$ is the charge of the $i^{\mathrm{th}}$ particle in the jet, and 
$p_i$ is the momentum. All charged particles with 
$p_\mathrm{T} > 0.5$~GeV, $|\eta|<2.5$ around the reconstructed B-meson were 
included in the jet, if the distance between the particle and the B-meson, 
$\Delta R = \sqrt{\eta^2+\phi^2}$, was less than 0.8. Particles not 
originating from near the primary vertex were excluded by requiring that the 
transverse impact parameter $d_0$ was less than 1 cm and the $z$-coordinate of 
the particle at the point of closest approach was within 5 cm of the $z$ of the 
primary vertex. The particles from the B-meson decay itself were excluded as 
well. In the analysis, the reconstructed B-meson was defined as 
$\mathrm{B_s}$ ($\mathrm{\bar{B}_s}$) if the jet-charge had 
$Q_{\mathrm{jet}}>+c$ ($Q_{\mathrm{jet}}<-c$), where $c$ is a tunable cut. 

The exponent $k$ was optimized to maximize the tagging quality factor 
$Q=D_{\mathrm{tag}}^2 \cdot \epsilon_{\mathrm{tag}}$, where $D_{\mathrm{tag}}$ 
is the tagging dilution factor, originating from wrong-sign tags 
($D_{\mathrm{tag}}=1-2W$, where $W$ is the fraction of wrong-sign tags), and 
$\epsilon_{\mathrm{tag}}$ is the tagging efficiency. The optimum parameters 
were found to be $k=0.2$, $c=0.2$, resulting in a quality factor of 3.85\% 
with $D_{\mathrm{tag}}=0.26$, $\epsilon_{\mathrm{tag}}=0.57$. 

The observable asymmetry is :
\begin{eqnarray*}
a_\mathrm{obs}(t) & = & D a_{CP}(t) = D \sin \phi_M \sin \Delta m_s t,
\end{eqnarray*}
where $D = D_{\mathrm{tag}} D_{\mathrm {back}}$ combines the experimental 
dilution factors due to mistagging and background. 
$D_{\mathrm {back}} = N_{\mathrm S}^{\mathrm {obs}}/( N_{\mathrm S}^{\mathrm 
{obs}} + N_{\mathrm{back}}^{\mathrm {obs}} )$ is the ratio of the number of observed 
signal events to the total number of observed events.

Assuming that the decay time resolution is $\Delta \tau$ =0.073~ps, the 
error on the CP asymmetry was estimated to be:
\begin{eqnarray*}
\delta(\sin \phi_M) & = & 0.27, \, \, x_s = \Delta m_s/\Gamma_s = 19,\\
\delta(\sin \phi_M) & = & 0.31, \, \, x_s = 30,
\end{eqnarray*}
where
\begin{eqnarray*}
\phi_M =  -2\lambda^2 \eta = -2\lambda \sin\gamma |V_{ub}|/ |V_{cb}|,
\end{eqnarray*}
where $\lambda=\sin\theta_C$, $\theta_C$ is the Cabibbo angle, $\eta$ is the 
height of the Unitarity Triangle and $\gamma$ is one of the angles of the 
Unitarity Triangle. 

\section{The $\mathrm{B_{s} \rightarrow J/\psi\phi}$ decay channel}

The $\mathrm{B_{s} \rightarrow J/\psi\phi}$ mode has a clean experimental
signature. For its reconstruction, the same approach as for the 
$\mathrm{B_{s,d} \rightarrow J/\psi\eta}$ was used. Details are given 
in~\cite{tdr2}

The $\mathrm{B_{s}\rightarrow J/\psi \phi}$ leads to three final state 
helicity configurations and their linear combinations are CP eigenstates 
with different CP parities~\cite{dighe}. For the extraction of the 
CP-violating weak mixing phase 
$\mathrm{\phi_{s}=arg(V_{cs}^{*}V_{cb}/V_{cs}V_{cb}^{*})}$ the helicity 
amplitudes need to be separated. Experimentally, one can measure three 
independent angles and the $\mathrm{B_{s}}$ proper decay time. The initial 
$\mathrm{B_{s}}$ flavour can be tagged in part of the events. The ATLAS 
precision for these measurements was determined by detector response simulations 
and was used as input to angular analyses based on a maximum likelihood fit in 
repeated Monte Carlo experiments. The likelihood function $\mathcal{L}$ is 
defined as:
\begin{eqnarray*}
\mathrm{\mathcal{L} =
\prod_{i=1}^{N}\frac{\int_{0}^{\infty}\mathcal{W}(t_{i},\Omega_{i})\cdot
Res(t,t_{i})}{\int_{t_{min}}^{\infty}\int_{0}^{\infty}\mathcal{W}(t.\Omega)\cdot
Res(t^{'},t)~dt^{'}~dt}}  
\end{eqnarray*}
with:
\begin{eqnarray}
\mathcal{W}(t,\Omega)=\epsilon_{1}\omega^{+}(t,\Omega)+\epsilon_{2}\omega^{-}(t,\Omega)+be^{-\Gamma_{0}t}
\end{eqnarray}
where $\epsilon_{1}=\epsilon_{2}$=0.5 for untagged events, $\epsilon_{1}$=1-r, 
$\epsilon_{2}$=r if the $\mathrm{B_{s}}$ is tagged as a particle and 
$\epsilon_{1}$=r, $\epsilon_{2}$=1-r if the $\mathrm{B_{s}}$ tagged as an 
anti-particle, r is the wrong tag fraction, b is the level of background and 
$\Gamma_{0}$ is the average decay rate of the background. The probability 
density $\omega^{\pm}$ is defined by:
\begin{eqnarray*}
\mathrm{\omega^{+(-)}=\frac{1}{(4\pi)^{2}}\frac{9}{8}\sum^{6}_{i=1} f(t)^{+(-)}_{i} 
F^{'}_{i}(\theta_{1},\theta_{2},\phi)} 
\end{eqnarray*}
 
in which $\mathrm{f(t)^{+(-)}_{i}}$ are expressed in terms of four parameters 
of $\mathrm{B_{s}}$ weak decay and oscillations: $\Delta \Gamma_{s}$, $\Gamma$, 
$x_{s}=\Delta m_{s}\Gamma$, $\xi$, and four independent parameters of helicity 
amplitudes:~$|A_{\Arrowvert}(t=0)|$, $|A_{\bot}(t=0)|$, and the two strong phases 
$\delta_{1}$ and $\delta_{2}$. The angular functions $\mathrm{F_{i}^{'}}$ are 
given in~\cite{maria1,maria2}, where more details about the analysis are also given. 
In the maximum likelihood fit the $\Delta \Gamma_{s}$, the $\Gamma$ and the weak phase 
$\phi_{s}$ were simultaneously determined along with two helicity amplitude 
values and their strong phases. The mixing parameter $x_{s}$ was assumed to 
have been measured in the $\mathrm{B_{s} \rightarrow D_{s}\pi}$ and 
$\mathrm{B_{s} \rightarrow D_{s} a_{1}}$ and was fixed. 

\begin{figure}
\hbox to\hsize{\hss
\includegraphics[width=\hsize]{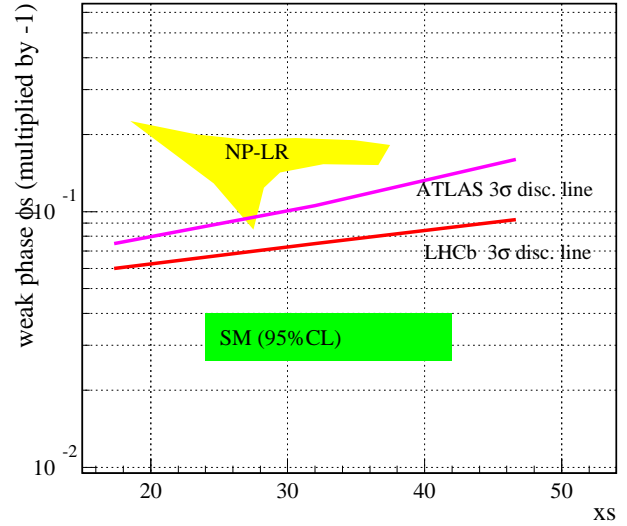}
\hss}
\caption{$\mathrm{\phi_s - x_s}$ regions allowed by the Standard Model
predictions and predictions from Left-Right Symmetric models (NP-LR), together
with the sensitivity lines for both ATLAS and LHCb.}
\label{fig:fig-sm}
\end{figure}

While all the eight parameters are independent in the theoretical models, 
the experimental 
resolution causes some of them become correlated, particularly the two strong 
phases, which prevented their simultaneous determination. Thus in the final 
analysis these two parametres were fixed. For an integral luminosity of 30 
fb$^{-1}$, $\Delta \Gamma_{s}$ can be detrmined with a relative error of 
12$\%$, while the precision of $\phi_{s}$ depends on the value of $x_{s}$ and 
on the proper time resolution, as shown in Fig.~3, where the 
discovery lines for ATLAS and LHCb are displayed in the 
($x_{s} - \phi_{s}$) plane, together with regions allowed by the Standard 
Model and by one example of new physics known as the Left-Right symmetric 
model~\cite{ball}. 

\section{Conclusions} 

ATLAS is expected to measure several parameters in the Bs meson system, which
will be valuable input for the flavour dynamics in the Standard Model and its
possible extensions.

A clear signal of the decay mode $\mathrm{B_{s} \rightarrow J/\psi\eta}$ 
is expected to be reconstructed with the ATLAS detector, and several 
measurements are possible such as the measurement of the branching fraction and
the Bs meson lifetime. The experimental resolution of the CP asymmetry measured
shows some sensitivity to very large asymmetries as predicted by some models
beyond the Standard Model. The measurement of the $\gamma$ angle seems, 
however to be compromised because the 
$\mathrm{B_{d} \rightarrow J/\psi\eta}$ is not observable. After 3 years at 
luminosity $\mathrm{10^{33}cm^{-2}s^{-1}}$ ATLAS is expected to determine 
$\Delta \Gamma_{s}$ with the relative errors of 12$\%$. The ATLAS 
discovery line (Fig.~3) shows that the precision is high enough 
to be sensitive to new physics.

\section{Acknowledgements}

This work has been performed within the ATLAS Collaboration, and the author
would like to thank all members of the collaboration, particularly members of 
the ATLAS B-physics working group.

\end{document}